\documentclass{bmcart}

\usepackage[utf8]{inputenc} 
\usepackage{graphicx}

\usepackage{amssymb}
\newcommand{\ket}[1]{\left| #1 \right>} 
\newcommand{\bra}[1]{\left< #1 \right|} 

\startlocaldefs
\endlocaldefs

\begin{document}

\begin{frontmatter}

\begin{fmbox}

\title{Quantum simulation of the Anderson Hamiltonian with an array of coupled nanoresonators: delocalization and thermalization effects}

\author[  
addressref={aff1},                   
]{\inits{JLV}\fnm{John} \snm{Lozada-Vera}}
\author[addressref={aff1},                   
]{\inits{AC}\fnm{Alejandro} \snm{Carrillo}}
\author[addressref={aff2},                   
]{\inits{OPSN}\fnm{Olimpio P} \snm{de S\'a Neto}}
\author[
   addressref={aff1},                   
  ]{\inits{JKM}\fnm{Jalil K} \snm{Moqadam}}
\author[
   addressref={aff3},                   
]{\inits{MDL}\fnm{Matthew D} \snm{LaHaye}}
\author[
   addressref={aff1},                   
   corref={aff1},
     email={marcos@ifi.unicamp.br}   
]{\inits{MCO}\fnm{Marcos C} \snm{de Oliveira}}
\address[id=aff1]{
 \orgname{Instituto de F\'\i sica ``Gleb Wataghin'', Universidade Estadual de Campinas}, 
  \postcode{13083-970}                                
  \city{Campinas - SP},                              
  \cny{Brazil}                                    
}
\address[id=aff2]{
 \orgname{Coordena\c c\~ao de Ci\^encia da Computa\c c\~ao,
Universidade Estadual do Piau\'i (UESPI)}, 
  \city{Parna\'iba-PI},                              
  \cny{Brazil}                                    
}
\address[id=aff3]{
 \orgname{Department of Physics, Syracuse University}, 
 \postcode{13244-1130}                        
  \city{Syracuse-NY},                              
  \cny{USA}                                    
}


\begin{artnotes}
\end{artnotes}

\end{fmbox}


\begin{abstractbox}
\begin{abstract}
The possibility of using nanoelectromechanical systems as a simulation tool for quantum many-body effects is explored.
It is demonstrated that an array of electrostatically coupled nanoresonators can effectively simulate the Bose-Hubbard
model without interactions, corresponding in the single-phonon regime to the Anderson tight-binding model.
Employing a density matrix formalism for the system coupled to a bosonic thermal bath, we study the interplay between
disorder and thermalization, focusing on the delocalization process. It is found that the phonon population remains
localized for a long time at low enough temperatures; with increasing temperatures the localization is rapidly lost due to thermal pumping of excitations into the array,
producing in the equilibrium a fully thermalized system. Finally, we consider a possible experimental design to
measure the phonon population in the array by means of a superconducting transmon qubit coupled to individual nanoresonators.
We also consider the possibility of using the proposed quantum simulator for realizing continuous-time quantum walks.
\end{abstract}
\begin{keyword}
\kwd{Quantum simulators}
\kwd{Nanoelectromechanical system}
\kwd{Anderson localization}
\end{keyword}


\end{abstractbox}
%

\end{frontmatter}

\textit{Introduction.}
The achievement of the ground state of mechanical motion using nanoelectromechanical systems (NEMS), as demonstrated
recently in remarkable experiments~\cite{ground_state1,ground_state2,ground_state3,Wollman952,PhysRevX.5.041037,PhysRevLett.115.243601}, opens up a new path for studying
quantum behavior in macroscopic systems.
Having been able also to coherently control and cleverly measure the state of the mechanical
resonator~\cite{ground_state1,wilson2015measurement,milburn2011introduction,suh2014mechanically,palomaki2013entangling,
lecocq2015resolving},
an immediate possibility to explore is the use of the NEMS as building blocks for fabricating analog quantum simulators
to reproduce many-body quantum physics~\cite{georgescu2014quantum,QS2}. Analog quantum simulators are dedicated and
controllable devices, which can imitate (within some accuracy) the evolution of certain types of Hamiltonians.
Various quantum systems have already been investigated for quantum simulation. Previously, quantum simulators were
experimentally implemented using ultracold quantum gases~\cite{bloch2012quantum},
trapped ions~\cite{blatt2012quantum}, photonic quantum systems~\cite{aspuru2012photonic} and superconducting
circuits~\cite{houck2012chip,schmidt2013circuit}. Recently, an array of optomechanical
resonators has been suggested for simulating many-body nonlinear driven dissipative quantum
dynamics~\cite{ludwig2013quantum}. 

One particularly important phenomena that emerges due to the wave-like nature of matter at the quantum regime is
Anderson localization~\cite{anderson1958absence,lee1985disordered,kramer1993localization}, a phenomena in which waves fail
to propagate in disordered media due to interference. Anderson localization has been observed with many experimental
setups, including
microwaves~\cite{dalichaouch1991microwave,chabanov2000statistical,chabanov2003breakdown},
light waves~\cite{wiersma1997localization,storzer2006observation,aegerter2006experimental,schwartz2007transport,
lahini2008anderson,sperling2013direct,segev2013anderson},
ultrasound~\cite{weaver1990anderson,hu2008localization} and
matter waves~\cite{billy2008direct,roati2008anderson,chabe2008experimental,kondov2011three,
jendrzejewski2012three,mcgehee2013three}.
Given the ability to achieve the quantum regime in mechanical resonators, it would be extremely appealing to observe
mechanical localization in an array of NEMS, where thermal effects become relevant.
Furthermore, arrays of NEMS could be used to simulate continuous-time quantum walk (CTQW) dynamics~\cite{farhi1998quantum}.
Quantum walks are the quantum version of random walks that have a crucial role in designing efficient quantum algorithms
that outperform classical algorithms~\cite{Portugal2013}. Implementation of CTQW has been already realized in a four-site
circle using a two-qubit nuclear magnetic resonance quantum computer~\cite{du2003experimental} and in a waveguide
array~\cite{perets2008realization}. A NEMS-based quantum simulator would provide a means to implement CTQW with phonons, 
opening up new opportunities for quantum algorithms and universal
quantum computation~\cite{childs2009universal}.

In this paper, we propose a $1$D array of coupled nanomechanical resonators for simulating the Anderson Hamiltonian, namely, a discrete
tight-binding model without on-site interactions. The coupling between resonators is electrostatic (capacitive coupling),
but could also be mechanic (elastic coupling) in order to be improved. The disorder can be induced in a controlled and predetermined manner through the appropriate variation in the design and nanofabrication of the geometrical dimensions of the nanoresonators in the array \cite{RevModPhys.86.1391}. A qubit coupled to
the chain can be used as a device for both, initializing and measuring the occupation probabilities of excitations. We also
discuss the physical implementations of CTQWs using the proposed quantum simulator.
\begin{figure}[h!]
\centering
\includegraphics[width=8cm]{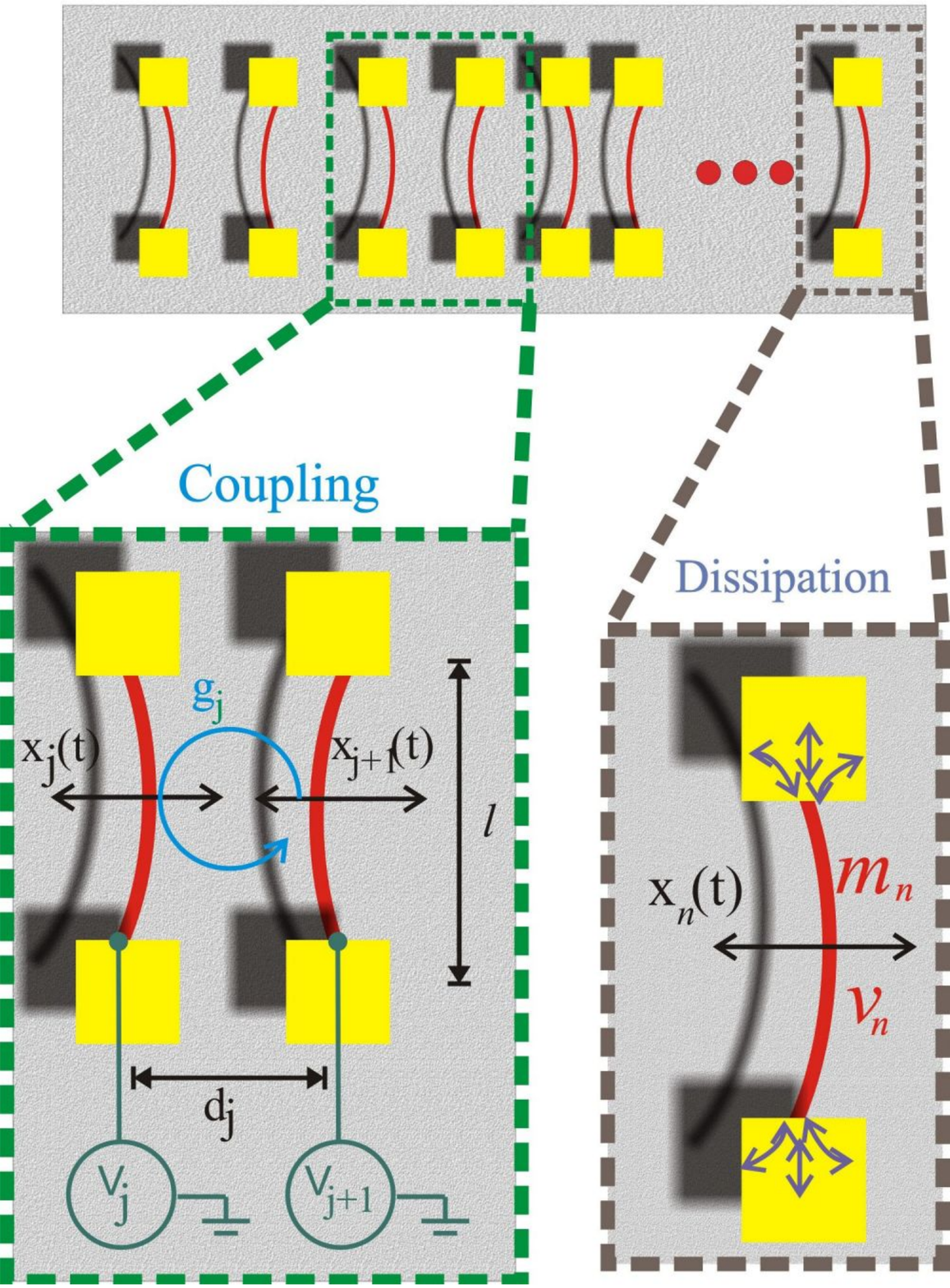} 
\caption{\csentence{A chain of electrostatically coupled mechanical resonators.} Every single resonator is charged by an individual
source, to induce electric potential diferences, and is considered as non-ideal by being coupled to individual thermal reservoirs.
Imperfections in the resonators fabrication naturally introduce a disorder in their respective principal oscillation mode,
hence, Anderson localization of phonon excitations can be observed for a sufficiently large array and sufficiently low
temperature. Depicted quantities are discussed within the text.}
\label{fig:Modelo-de-cadeia}
\end{figure}

\section{Model}
An array of $N$ capacitively coupled electromechanical resonators, as depicted in Fig.~\ref{fig:Modelo-de-cadeia}~(upper panel), can be described by the Hamiltonian
\begin{equation}
\label{H1}
 \mathcal{H}  =  \sum_{j=1}^{N}\left(\frac{p_{j}^{2}}{2m_{j}}+\frac{1}{2}m_{j}\nu_{j}^{2}x_{j}^{2}\right)
               + \sum_{j=1}^{N-1}U_{j},
\end{equation}
where $x_{j}$, $p_{j}$, $m_{j}$ and $\nu_{j}$ are the displacement from the equilibrium position, momentum, mass and
frequency, respectively, associated with a single mechanical mode of each resonator.  $U_{j}$ is the
electrostatic interaction energy between the pair of nearest-neighbor resonators $j$ and $j+1$. The interaction between
a resonator and its second nearest neighbor is small enough to be disregarded.
Using a simple parallel plate capacitor model~[see Fig.~\ref{fig:Modelo-de-cadeia}~(coupling)], the energy $U_j$ in
Eq.~(\ref{H1}) can be written as
\begin{equation}
\label{eq:electrostatic_V}
 U_{j}=\frac{1}{2}C_{j}\Delta V_{j}^{2}=\frac{1}{2} \frac{ \epsilon_{0} A \Delta V_{j}^{2} }{d+[x_{j}(t)-x_{j+1}(t)]},
\end{equation}
where $C_{j}$ and $\Delta V_{j}=\left|V_{j+1}-V_{j}\right|$ are the capacitance and the voltage difference between the
resonators $j$ and $j+1$, respectively. The capacitance $C_{j}$ is expressed in terms of the vacuum permittivity
$\epsilon_{0}$, the resonator area $A$ and the equilibrium center-of-mass separation $d$. Note that the inevitable disorder appearing in $d$ is here neglected because its effect would be to introduce a disorder in the coupling energies, which are too small to be taken into account. The oscillation amplitudes are close to the zero point fluctuations, hence,
$\left|\delta x_j(t)\right|=\left|x_{j}(t)-x_{j+1}(t)\right|\ll \nobreak d$ and the electrostatic potential~(\ref{eq:electrostatic_V}) can be expanded in powers of
$\delta x_j(t)/d$, which up to the second order gives
\[
\label{eq:Uapprox}
 U_j \simeq \frac{\epsilon_{0}A \Delta {V_j}^2}{2d} \bigg[ 1 - \frac{\delta x_j(t)}{d} 
      + \frac{\left[\delta x_j(t) \right]^2 }{2d^2} \bigg]. \nonumber
\]
The linear terms will cancel out in the summation of $U_j$, and only remaining linear terms are $(x_1(t)-x_N(t))/{d}$, which nonetheless will also be negligible in comparison to the remaining second order terms, due to the rapid oscillation of $x_j(t)$. We now quantize the system, by replacing the classical variables with the operators written in terms of the annihilation
and creation operators of the resonator excitation, $a_{j}$ and $a_{j}^{\dagger}$. In the interaction picture, the
position and momentum operators are given by
$\hat{x}_{j}(t)=\sqrt{\hbar/(2m_{j}\nu_{j})}(a_{j}^{\dagger}e^{i\nu_{j}t}+a_{j}e^{-i\nu_{j}t})$
and $\hat{p}_{j}=i\sqrt{\hbar m_{j}\nu_{j}/2}(a_{j}^{\dagger}e^{i\nu_{j}t}-a_{j}e^{-i\nu_{j}t})$, respectively.
Substituting them in Eq.~\ref{eq:Uapprox}, the rapidly oscillating terms vanish in the rotating wave approximation (RWA) and
Hamiltonian~(\ref{H1}) is converted to 
\begin{equation}
\mathcal{H} = \sum_{j=1}^{N}\hbar\omega_{j}a_{j}^{\dagger}a_{j}- \sum_{j=1}^{N-1}\hbar g_{j}(a_{j}^{\dagger}a_{j+1}
+a_{j}a_{j+1}^{\dagger}),
\label{eq:A_H}
\end{equation}
where $ g_j = \epsilon_0 A \Delta {V_j}^{2} x^{\mathrm{zpf}}_jx^{\mathrm{zpf}}_{j+1} / 2 d^3 $ is the tunnelling energy,
$x^{\mathrm{zpf}}_j=\sqrt{\hbar/(2m_{j}\nu_{j})}$ is the zero point fluctuation of the resonator $j$ and $\omega_j=\nu_j+\frac{\epsilon_0 A \Delta V_{j}^{2}}{2 d^3 m_j \nu_j}$ is the
rescaled mode frequency.
$\omega_j$  is randomly distributed over the
range $[\overline{\omega}-\Delta,\overline{\omega}+\Delta]$, where $\overline{\omega}$ is the average frequency  of the resonators and
$\Delta$ is called the disorder intensity.
We assume that the difference between any pair of disordered mode frequencies is small, hence the resonance condition is
almost satisfied. In other words $\Delta\ll\overline{\omega}$. In fact for all simulations considered here we have assumed $\Delta\le 10^{-2} \overline{\omega}$. Moreover  in any practical situation  $\overline{\omega}\gg g_j$, but $\Delta \ge g_j$ (See  Section 4 for discussion of engineering the system parameters, and Table 1 for experimental values), and therefore the validity of the RWA is always assured. Note that the coupling energies $g_{j}$ depend on several experimental parameters
related to the resonator fabrication such as mass, frequency and geometry, as well as the chain parameters such as the
distance between the resonators and the potential difference. Therefore they could also be random variables
that would define an off-diagonal disorder. However, since $\omega_j\gg g_j$, the disorder in $g_j$ is negligible in comparison to the diagonal one introduced by $\Delta$. To simplify the calculations we fix all the
tunneling energies to $g_i=J$.

The Hamiltonian (\ref{eq:A_H}) is the Bose-Hubbard Hamiltonian without inter-mode interactions, and a diagonal disorder in the on-site energies $\omega_j$. When the total number of phonons is restricted to $1$, this Hamiltonian reduces to the Anderson tight-binding Hamiltonian (see below).

\section{Closed System: Anderson Localization} 
Localization, from a general point of view, is a mesoscopic phenomena displayed by waves as they propagate through a disordered medium. It is built upon the interference of the many randomly scattered waves, which at sufficiently large distances, collude to produce a suppression in the amplitude of propagation. This behavior is characterized by the exponential decay of the wave-function, with a decay length $\xi$, known as the localization length. The celebrated scaling theory of localization \cite{Gang_of_4} describes how the transition between the different diffusion regimes depend upon the size $L$, and dimension $D$, of the system, independent of the microscopic intricacies of the disorder. According to the theory, depending on $L$, three different transport regimes can be recognized: ballistic ($\xi\gg L$), where the size of the system is too small and scattering events are rare; diffusive ($\xi\lesssim L$), where some weak localization effects may take place; and strong localization ($\xi\ll L$), for large systems \cite{MD09}. Of course, the detailed form of the localization depends on the type of disorder potential considered and its energy spectrum.

In the Anderson model (which is the one of interest in this paper), disorder is modeled by a $\delta$-correlated potential $V$, consisting of a series of spatially uncorrelated barriers\footnote{For instance in 1D, this means that the correlation function of the potential $g(z)=\langle V(x)V(x+z) \rangle = \Delta^2 \delta(z)$}, with a finite maximum amplitude intensity $\Delta$, called disorder strength. One can as well picture a periodic lattice with randomly shuffled on-site energies. Strictly speaking, a proper phase transition from extended to localized states only exists for $D=3$, at some critical disorder intensity or, interchangeably, some critical energy known as the mobility edge. For $D<2$ it can be shown that all states are localized, but the localization length for $D=2$ can easily exceed the size of the system for weak disorder or high enough particle energies, thus admitting the existence of a diffusive transport regime. In contrast, for $D=1$, (almost\footnote{Almost here means for almost every realization of the potential and for almost every energy, except perhaps for some pathological potentials. See \cite{vanTiggelen} for example for more rigorous definitions of localization.}) all single-particle eigenstates are localized even for a vanishingly small $\Delta$ and there is no phase transition\footnote{It is interesting to mention anyway that for some long-range correlated disorders an effective mobility edge can appear even in the 1D case \cite{IK99}.}.

\begin{figure}[h!]
\begin{centering}
\includegraphics[width=8cm]{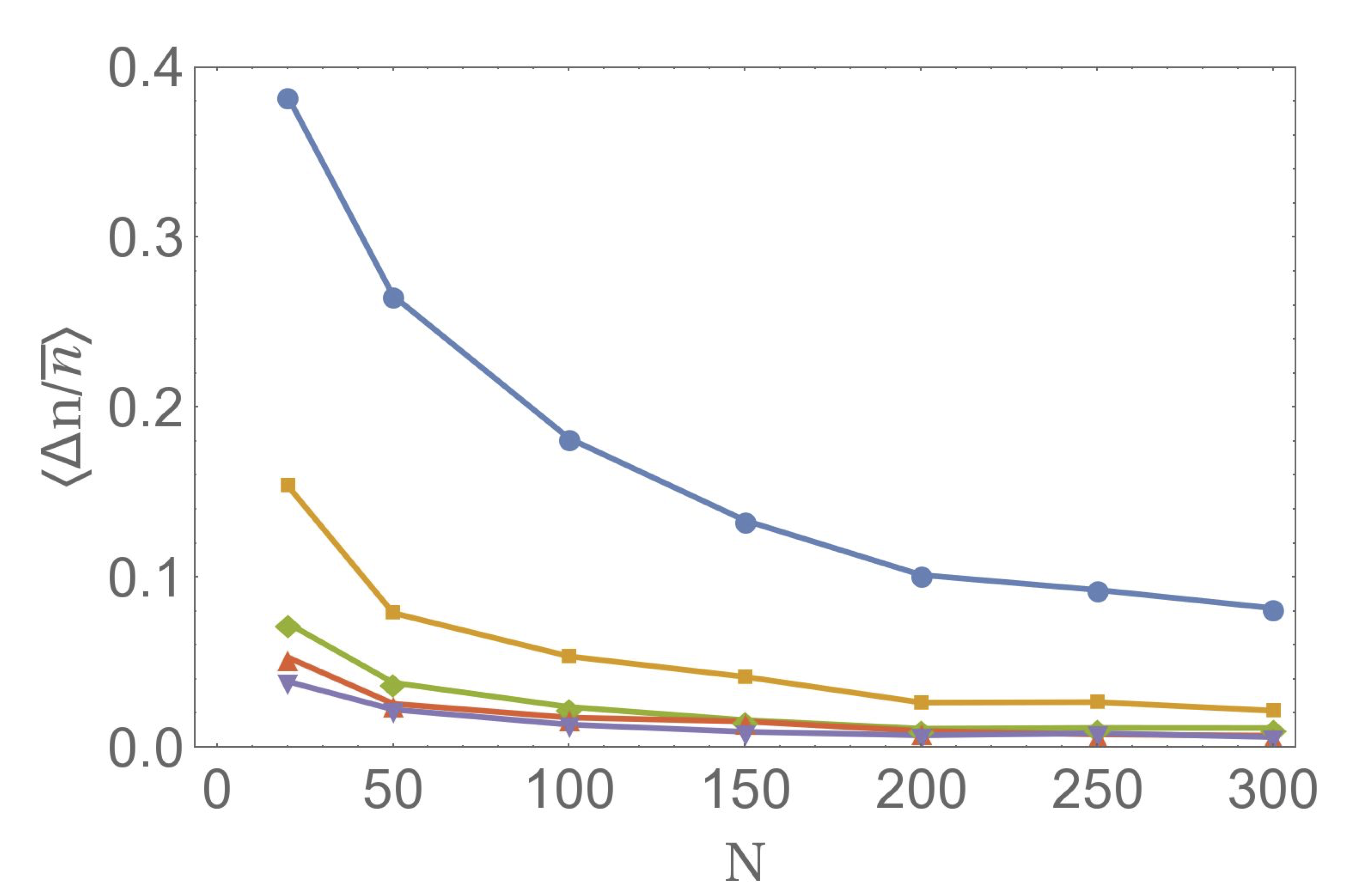}
\par\end{centering}
\caption{\csentence{Population second moment.} Average over 500 disorder realizations of the relative second moment in the population distribution $\Delta n/n_{av}$ for the ground state of the Anderson Hamiltonian, as a function of the system size N and for different disorder
intensities (from top to bottom $\Delta/J=2,5,10,15,20$) showing the convergence of the dispersion of the population as the number of sites is increased.}\label{fig2}
\end{figure}

For the description of the linear chain of resonators, we can disregard the free propagation between lattice sites, focusing only on the occupation amplitudes of each site \footnote{Discrete models bring essentially the same qualitative results as continuous ones regarding localization effects, and they are solvable for many disorder distributions.}. In the single-phonon situation, Hamiltonian (\ref{eq:A_H}) turns into ($\hbar=1$):
\begin{equation}
\mathcal{H}_A = \sum_{j=1}^{N}\omega_{j} \ket{j}\bra{j}+  J(\ket{j}\bra{j+1}+\ket{j+1}\bra{j}),
\label{eq:H_And}
\end{equation}
where the states $|j\rangle$ are the occupation amplitudes of each resonator, $J$ is the coupling rate between neighbouring sites, and the set of $\omega_{j}$ are assumed to follow a uniform random distribution around the mean value $\omega$. For this model, numerical simulations are easy to perform, for example using closed boundary conditions at both edges of the chain, and diagonalizing Hamiltonian (\ref{eq:H_And}) directly, varying the total number of sites $N$ and disorder intensity $\Delta$, for many realizations of the disordered potential\footnote{In this work all the simulations were performed for $500$ realizations of the disorder.}. To set the stage for further discussions, we start in fig. (\ref{fig2}) by plotting the behavior of the averaged relative dispersion in the site population for the ground state of the resonators $\ket{\psi^{0}}$, given by $\langle\langle\Delta n/n_{av}\rangle\rangle = \langle\langle[\sum_{j}^{N}j^2 |\psi_{j}^{(0)}|^2-n_{av}^2]^{1/2}/n_{av}\rangle\rangle$, as a function of the system size ($N$ ranging between $1$ to $300$ sites) and for different disorder strengths ($\Delta/J=2,5,10,15,20$). Here $n_{av}=\sum_{j}^{N}j |\psi_{j}^{(0)}|^2$, is simply the mean value of the site population, and $\psi_{j}^{(0)} = \langle j|\psi^{0}\rangle$ are the different population amplitudes for the ground-state; the double angular brackets denote the average over the disorder realizations. It can be noted the sensibility of the model for small sizes, since although the eigenstates are always localized, the relative dispersion in population sites only converges for a large enough system, a situation that is more relevant as the disorder gets weaker as can be seen comparing the different curves for the region $N\lesssim 50$. This dependence on the system size is clear in the curve in fig.(\ref{fig2}) for $\Delta/J=2$. It would be required a much larger system for the curve to converge closer to the other ones.

\begin{figure}[h!]
\includegraphics[width=13cm]{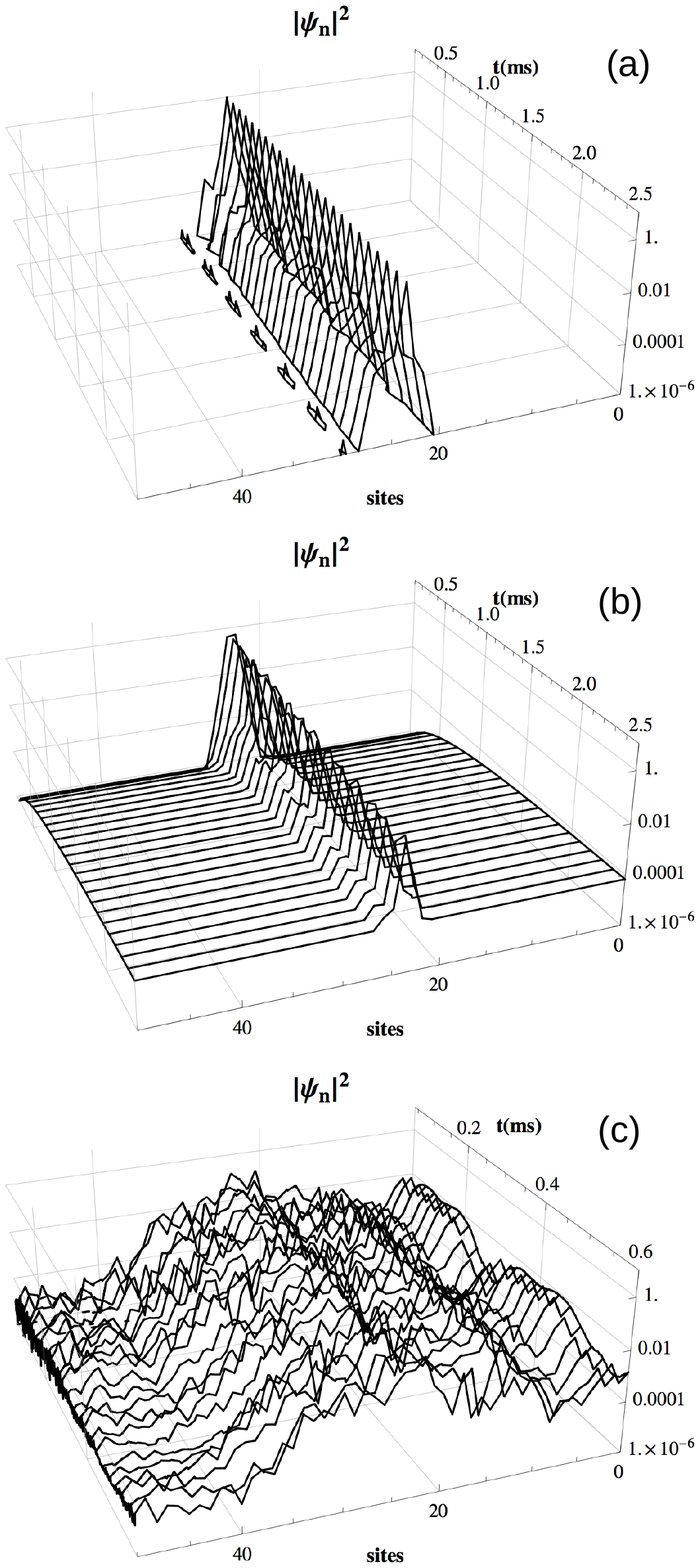}
\caption{\csentence{Density profiles.} Density profiles of the phonon population for a  chain of 51 nanomechanical resonators. (a) Depicts the situation for a closed system with $\Delta/J = 15$, while (b) and (c) for open systems affected by thermal reservoirs  with $\bar{n}= 10^{-2}$ and $\gamma =1$ MHz. In (b) $\Delta/J = 15$; in
(c) $\Delta/J = 2$.}\label{watf}
\end{figure}

The typical exponential profile for the population density is plotted in Fig.~\ref{watf}(a) for  $\Delta/J=15$. A single excitation in the central site for a chain of 51 resonators is assumed as the initial condition. In a Bose-Einstein Condensate (BEC), the measured quantity of interest  is the density of states at a given
position in the
lattice, namely, $\left|\psi(x)\right|^{2}$~\cite{billy2008direct,roati2008anderson,chabe2008experimental,kondov2011three,
jendrzejewski2012three,mcgehee2013three}. In our simulator, the quantity to be measured is the population of the first excited
state of a given resonator $j$, which is $\rho_{\ket{\omega_{j}}\bra{\omega_{j}}}$. Given the discrete nature of our system,
we cannot expect a smooth Gaussian-to-exponential transition in the \textit{population} profile. However, we see that due to
the presence of disorder in the diagonal terms of the Hamiltonian, $\omega_j$ in~Eq. (\ref{eq:A_H}), the spatial distribution of the excitations present
in the chain remain always close to the initial spatial distribution. Note that in Fig. \ref{watf} the initial nonstationary regime is not shown. This short time which is disregarded corresponds to the transitory path to equilibrium.

It is interesting to explore the physics that is difficult to be included in the Anderson model, and which is nonetheless
present in a real implementation of the system. Specifically, the influence of thermal effects due to a surrounding reservoir
for the resonators is considered in the following.

\section{Open System} Here, we write a master equation for the Anderson Hamiltonian~(\ref{eq:A_H}) to describe the
effects of the environment on the chain and consequently to investigate the corresponding influence on the localization
of the states. In this way, we need to take into account phonons and weak electromagnetic fields that are surrounding the
nanoresonators. Assuming that each resonator in the chain is coupled to a bosonic bath, the corresponding interaction
Hamiltonian takes the form
\begin{equation}
\mathcal{H}_{\mathrm{I}} = -\sum_{j,l} \Gamma_{jl} \big(a_{j}^{\dagger}b_{j,l} + a_{j}b_{j,l}^{\dagger}\big),
\label{eq:BH_bathinteraction}
\end{equation}
where $b_{j,l} $ and $b_{j,l} ^{\dagger}$ are the $j$-th resonator bath annihilation and creation operators and $\Gamma_{jl}$ is the coupling between
resonator-bath modes. In order to derive the master equation it is supposed that each resonator is weakly coupled to its respective bath, and that all bosonic baths are in thermal equilibrium, with mean thermal phonon number $\overline{n}$. The master equation describing the dynamics of the chain is then
\begin{eqnarray}
\label{eq:ME}
\dot\rho_{S}(t) &=& -\frac{i}{\hbar} \Big[ \mathcal{H} ,\rho_{S}(t) \Big] \
  +\sum_{j}\frac{\gamma_{j}}{2}\left[\overline{n}\big(2a_{j}^{\dagger}\rho_{S}(t)a_{j}-a_{j}a_{j}^{\dagger}\rho_{S}(t)
 -\rho_{S}(t)a_{j}a_{j}^{\dagger}\big)\right.\nonumber \\
  &&\left.+(\overline{n}+1)\big(2a_{j}\rho_{S}(t)a_{j}^{\dagger}-\rho_{S}(t)a_{j}^{\dagger}a_{j}
 -a_{j}^{\dagger}a_{j}\rho_{S}(t)\big)\right],
\end{eqnarray}
where $\rho_{S}$ is the system density operator, $\gamma_{j}$ is the effective resonator-bath interaction strength
and $\overline{n}$ is the bath mean
thermal phonon number, which is specified by using the
Bose-Einstein statistics. We have considered $\gamma_{j}=\gamma= 1$ MHz for all simulations in this work. Remark that since
we are interested in considering the Anderson model for a single phonon population in the chain we must keep the temperature
reasonably low. Considering actual temperatures reached in experiments for mechanical resonators in the quantum regime,
this corresponds to very low $\overline{n}$.  We have considered $\overline{n}$ varying from $10^{-4}$ to
$10^{-1}$~\cite{ground_state1}, which nonetheless is sufficient to see a disturbance in the localization mechanism, without compromising our numerical simulation.

The effect of the reservoir on the system include loss of excitations due to imperfections in the medium.
It simply means that the single phonons in the chain, eventually would \textit{leak} into the medium. This phenomena can
be understood like an incoherent spontaneous emission. The expected behavior of a system subject to an amplitude damping
channel is the escape of all excitations in the chain.
However, while such decoherence process is taking effect, we can argue the existence of enough quantum correlations as to
assure the existence of localization phenomena. Working in a weak resonator-bath interaction regime (something like three
orders of magnitude less than the resonator-resonator interaction), the correlations remain in the system until the
\textit{leaking} process has taken away all measurable probability of excitations. We can clearly see this behavior in
Fig.~\ref{watf}(b) for $\gamma=1$ MHz, $\bar{n}=10^{-2}$, and normalized disorder $\Delta/J=15$. Similarly to Fig.~\ref{watf}(a) a single excitation in the central resonator is taken as the initial condition. We see after equilibration a behavior
which similar to he one characterising localization in Fig.~\ref{watf}(a) but for a increasing uniform thermal excitation base. In Fig.~\ref{watf}(c) the same initial condition and dynamics is employed but for a smaller
disorder $\Delta/J=2$. Now we see a typical thermal behavior of a non-localized density profile. Remark that for $\gamma=1$ MHz there is enough time for equilibration. 
\begin{figure}[h!]
\begin{centering}
\includegraphics[width=8cm]{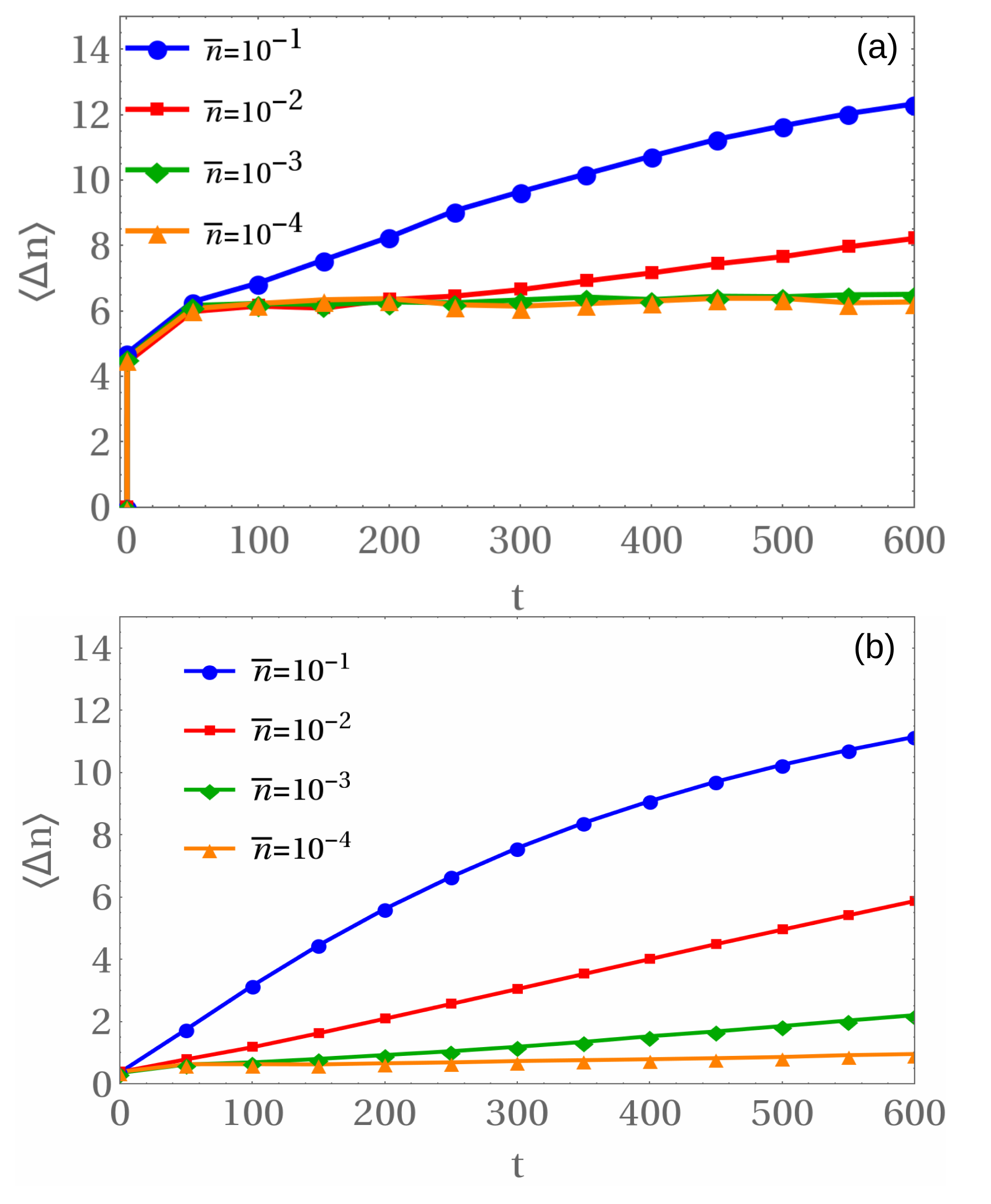} 
\par\end{centering}
\caption{\csentence{Phonon population dispersion.} Time evolution of the resonators phonon population dispersion for a  chain of 51 nanomechanical resonators. $\langle\langle \Delta n\rangle\rangle$, as a function of the thermal phonon number, $\bar n$, for (a) $\Delta/J=2$; and (b) $\Delta/J=15$. The higher $\bar{n}$ gets, the stronger will be the delocalization.}\label{loss}
\end{figure}

In Fig. 4 it is depicted more clearly the effect of the thermal excitation over the phonon localization. There it is plotted the time evolution of the resonators phonon population dispersion, $\langle\langle \Delta n\rangle\rangle$, as a function of the thermal phonon number, $\bar n$, for two different disorder strengths: 4(a) $\Delta/J=2$ and 4(b) $\Delta/J=15$, and for a single excitation  in the central resonator as initial condition. It can be seen that the higher the temperature (or $\bar{n}$) gets, the stronger will be the delocalizing effect as expected, very much independent of the disorder intensity. However for the lower temperatures (green and orange curves), the phonon population remains localized around the initial excited resonator, even in the presence of leaking in the total population of excitations due to dissipative effects. Furthermore, for the higher temperatures (blue and red), the localization is lost due to thermal pumping of excitations into the array, producing a fully thermalized state.

\begin{figure}[h!]
\begin{centering}
\includegraphics[width=8cm]{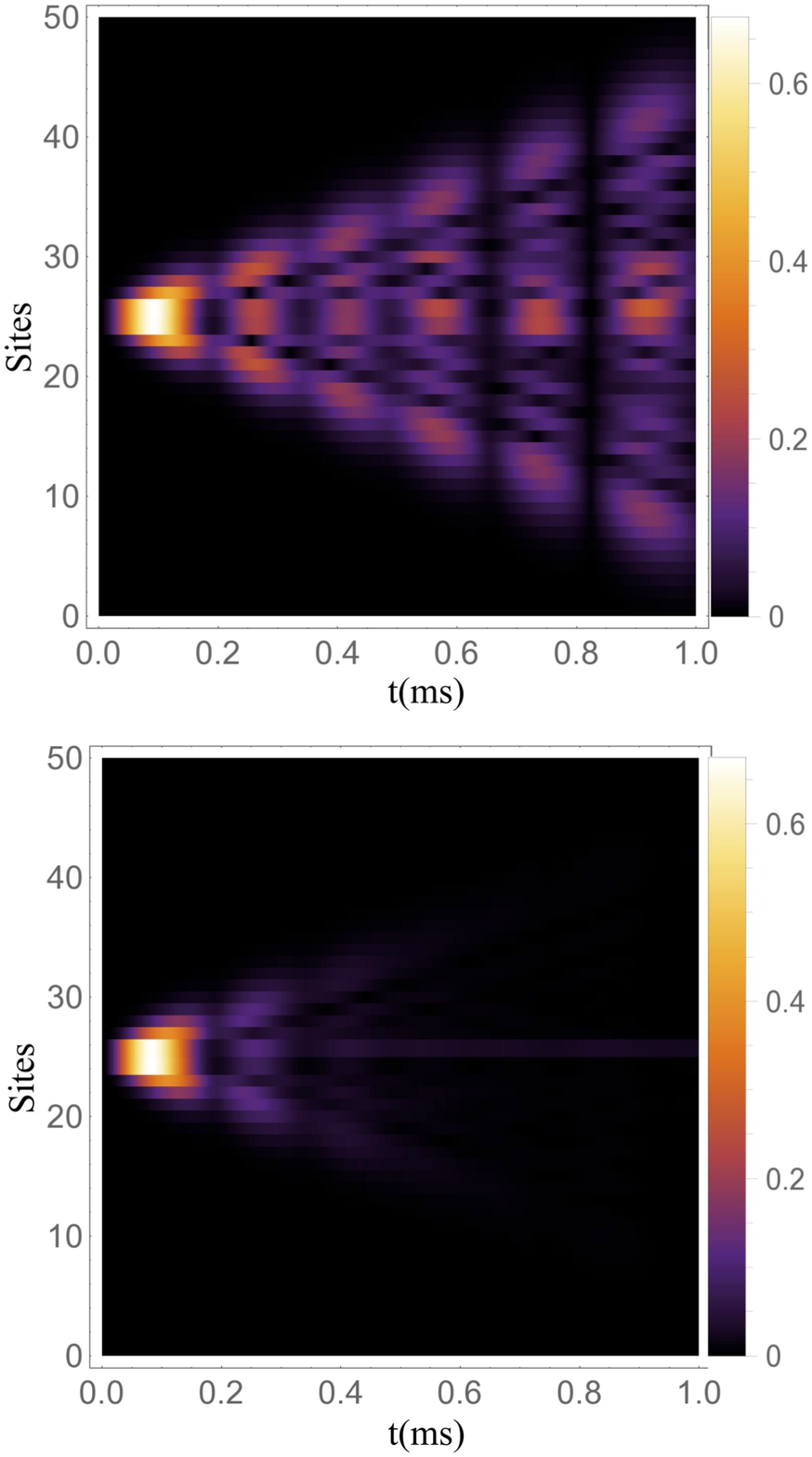}
\caption{\csentence{Concurrence density.} Concurrence density through the chain between the initially excited nanomechanical resonator (the central nanomechanical resonator) and all
the others with $\Delta/J=0$ (ballistic expansion). Upper panel: Closed system characterized by $\gamma=0$ and $T=0$.
Lower panel: open system with $\gamma/J=0.05$, and the mean number of phonons $\bar{n}=10^{-2}$.}
\label{fig:concurrencefree}
\par\end{centering}
\end{figure}

\begin{figure}[h!]
\begin{centering}
\includegraphics[width=8cm]{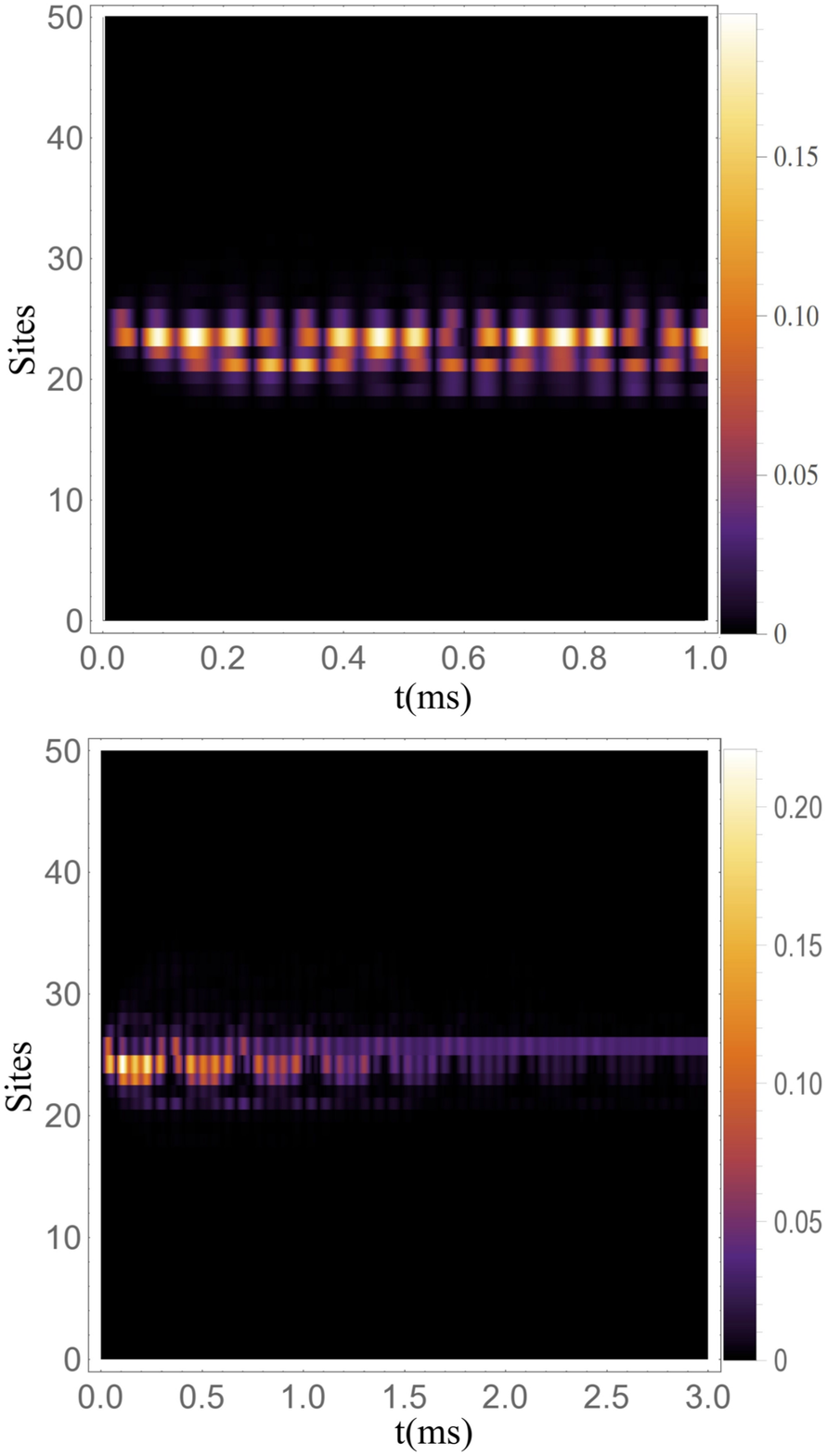}
\caption{\csentence{Concurrence density.} Concurrence density through the chain between the initially excited nanomechanical resonator (the central nanomechanical resonator) and all the
others with $\Delta/J=10$, that is the localized system. Upper panel: Closed system characterized by $\gamma=0$
and $T=0$. Lower panel: open system with $\gamma/J=0.05$, and a mean number of phonons $\bar{n}=10^{-2}$.}
\label{fig:concurrencedisorder}
\par\end{centering}
\end{figure}

With an initial condition available experimentally, namely, exciting a central nanomechanical resonator in the chain (see Section 4), the fast dynamics creates
correlations in nearby resonators. These correlations imply long-time entanglement between the resonators, which in turn give
us the possibility to maintain localization until thermalization is reached.
To investigate the correlations in a finite chain of nanoresonators we consider the concurrence which is a bipartite
entanglement measure and defined as~\cite{Wooters}:
\begin{equation}
 C(\rho)= \max \Big \{ 0,e_{1}-e_{2}-e_{3}-e_{4}\Big\},
\end{equation}
where $e_{i}$ are the square roots of the eigenvalues in decreasing order of the positive definite
matrix $\rho\tilde{\rho}$, where
\begin{equation}
 \tilde{\rho}= \Big ( \sigma_{y}\otimes \sigma_{y} \Big) \rho^{*}\Big ( \sigma_{y}\otimes \sigma_{y}\Big),
\end{equation}
in which $\rho^{*}$ is the complex conjugate of the density matrix $\rho$ and $\sigma_{y}$ is the $Y$ Pauli matrix.
Concurrence belongs to the interval $[0,1]$ where $C(\rho)=0,1$ corresponds to separable states and maximally entangled
states, respectively. we measure the concurrence between the initially excited nanomechanical resonator (at the centre of the chain) and each of
the other nanomechanical resonators under different conditions of the disorder and coupling with the environment. The results are shown in
Figs.~\ref{fig:concurrencefree} and \ref{fig:concurrencedisorder}.

Figures~\ref{fig:concurrencefree}~and~\ref{fig:concurrencedisorder} show that the behavior of the entanglement is
consistent with that of the population probability distribution from Fig.~\ref{watf}. The concurrence and the
population both spread through the available sites or become localized due to the disorder. The presence of an environment
does not introduce new dynamics. The only effect of the environment, as expected, is to destroy quantum coherences with a
rate proportional to $\exp(-\gamma t )$, regardless of the presence of localization.
The figures also show the \textit{sudden death} of the entanglement~\cite{Almeida} in which the concurrence between the
central nanomechanical resonator and all the others destroys periodically.

The inclusion of thermal effects does not immediately change the main feature of the localization (Fig. 6, bottom). The system is quickly localized as
expected from the Anderson-like Hamiltonian with disordered diagonal terms. With dissipation, localization is still a
\textit{quantum} effect in the sense that correlations between the resonators remain until the dissipation takes away all
possible dynamics. As time goes on, depending on the temperature, all the states become thermalized. Even for very
low temperatures, before all excitations leave the chain, the equilibrium state will be a thermalized state. However, the thermal relaxation rate is slow enough that localized phonon populations could still be measured prior to decaying.

\begin{figure}[h!]
\begin{centering}
\includegraphics[width=8cm]{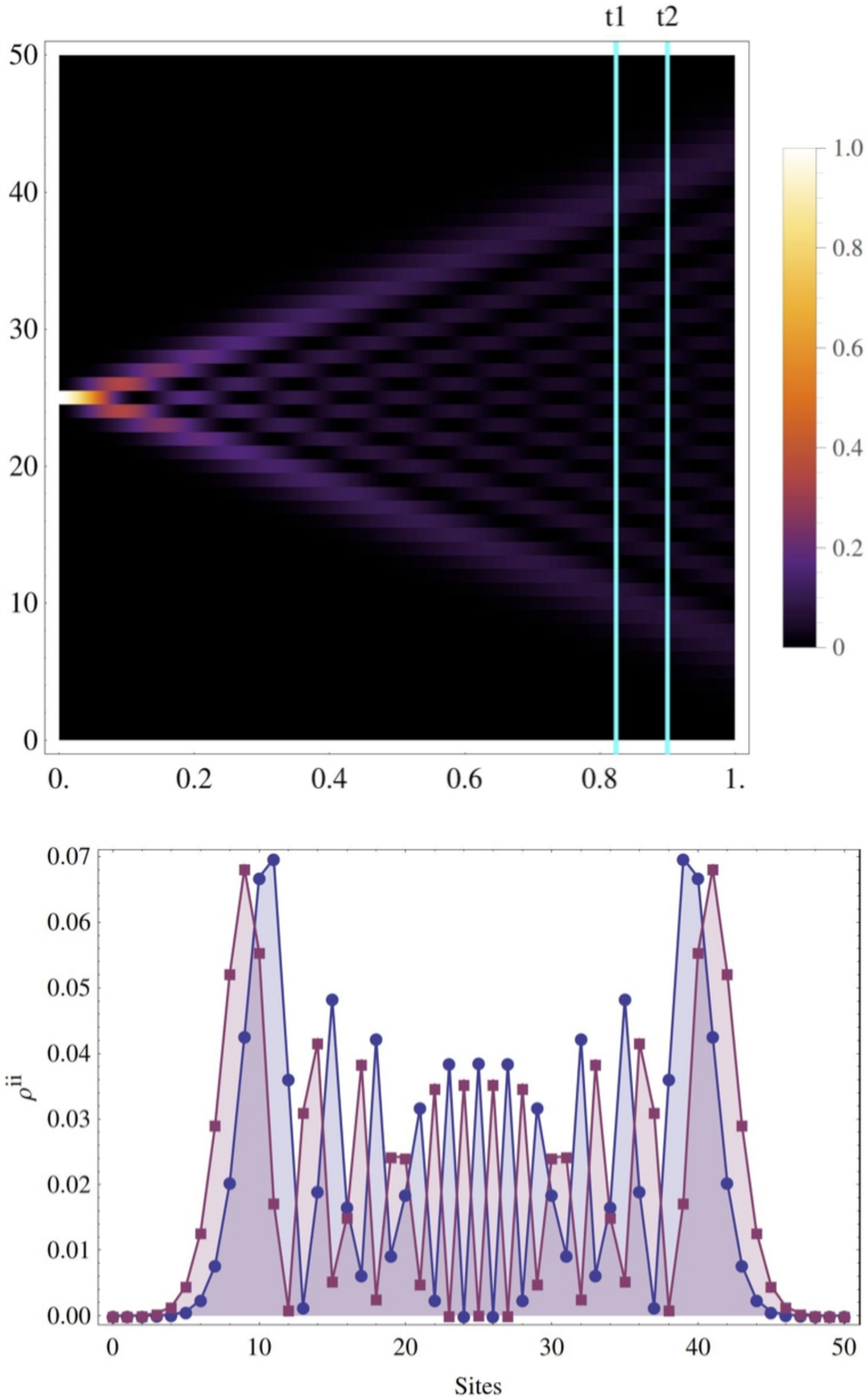}
\caption{\csentence{Phonon density probability.} Phonon density probability and density profile for two specific times. The profile show the typical behaviour of
the continuous quantum walk, which is implemented from the Anderson Hamiltonian. Parameters are the same as in the previous
figures.}\label{qwalk}
\par\end{centering}
\end{figure}

\textit{Continuous-Time Quantum Walk.}
The Anderson Hamiltonian given in Eq.~(\ref{eq:A_H}) also generates CTQW dynamics. For a small disorder, an initial phonon injected in the
central nanoresonator spreads ballistically and the standard deviation of the corresponding probability distribution over
the chain increases linearly with time. The measurement method which is described in the following
section can be used to reconstruct the probability distribution, shown in Fig.~\ref{qwalk}.
It is also possible to inject two or more phonons in the chain to investigate the
multi-particle quantum walks. That, specifically, provides a means for simulating bosonic particles and is going to be addressed elsewhere.

\section{Measurement} 
\begin{figure}[h!]
\begin{centering}
\includegraphics[width=8cm]{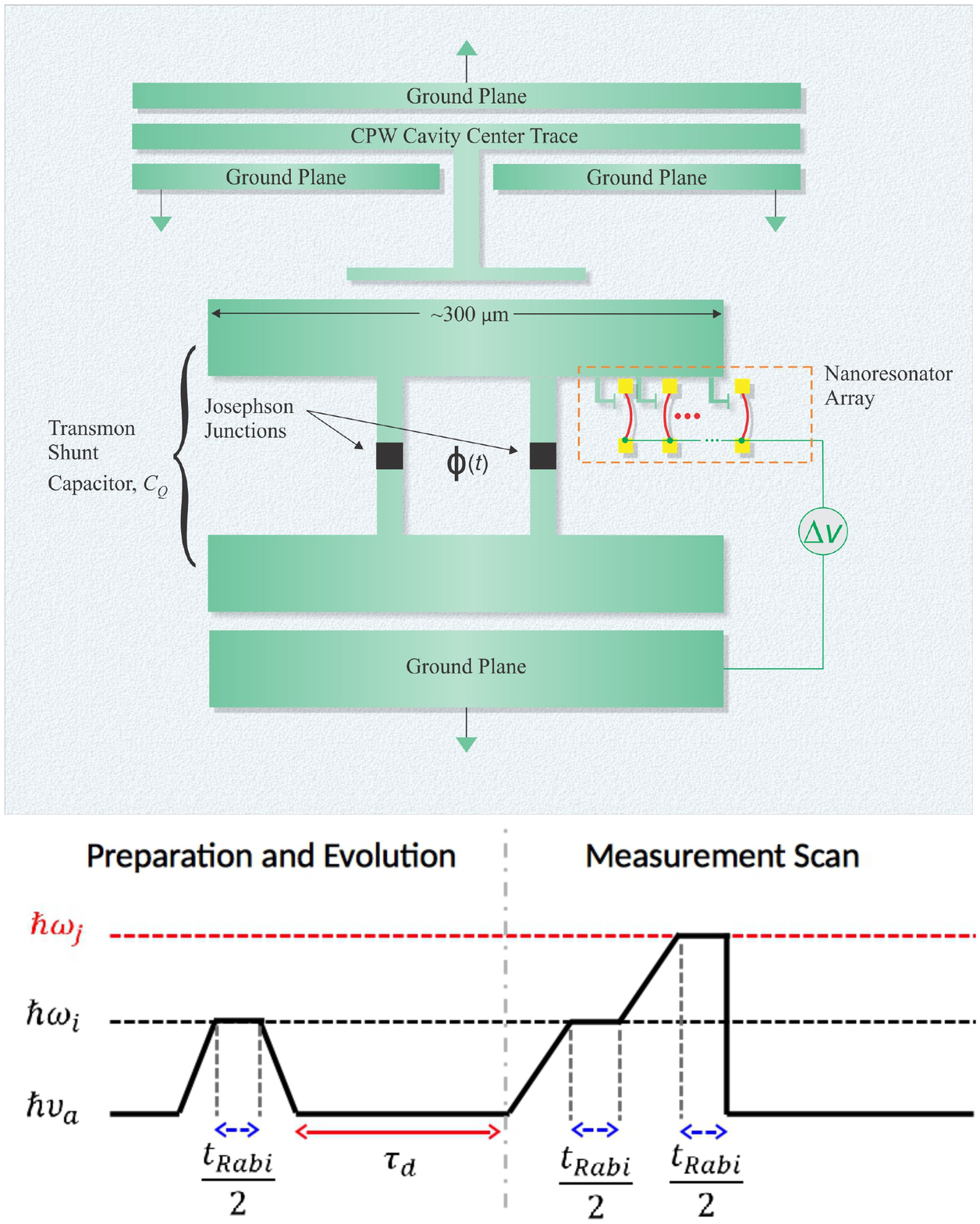}
\caption{\csentence{Illustration of the proposed layout and experimental protocol
for implementing the simulation of the Anderson Hamiltonian.} (a) Sample
layout. The voltage-biased nanoresonator array would be capacitively-coupled
to one pad of transmon's shunt capacitor $C_Q$. The transmon would also be
capacitively-coupled to a CPW cavity for read-out. Note that layout is not
drawn to scale - an array of 40 nanoresonators would extend only several
microns along the transmon pad. (b) Protocol for preparing an initial photon
excitation in a nanoresonator ($\omega_i$) and detecting the phonon after
delay time $\tau_d$ in a second nanoresonator ($\omega_j$). The transmon
transition energy $\nu_a$ would be controlled by application of flux
ramp $\Phi(t)$ to the transmon. Note the ramp rates as shown are not drawn
to scale.}\label{nmeas}
\par\end{centering}
\end{figure}

Quantum measurement is a crucial step in quantum simulation.  For our simulator, we envision using a single superconducting transmon qubit~\cite{koch2007charge,pirkkalainen2013hybrid} for the read-out of the phonon density profile in the nanoresonator array.  In this scenario, the nanoresonators would be metallized and coupled capacitively~\cite{pirkkalainen2013hybrid} to the transmon as shown in Fig. \ref{nmeas}a.
It is important that the location of the nanoresonators be determined. It can be done by mapping the nanoresonator frequencies in the array, which could be compiled either before or after the measurements using electron beam imaging (to determine with high precision the nanoresonators' geometries) and finite element simulations.  Through a filtered circuit connection~\cite{hao2014development}, a DC voltage V could be applied to establish the interaction between each individual nanoresonator and the transmon, which can be modeled using the Jaynes-Cummings Hamiltonian~\cite{ground_state1,irish2003quantum} – note the same applied voltage would serve to couple the nanoresonators to each other. The full measurement Hamiltonian, incorporating the entire array of nanoresonators, would thus be given by
\[
  \tilde{\mathcal{H}} = \frac{1}{2}\hbar\nu_{a}\sigma_{z} + \hbar\sum_{j}\omega_{j}a_{j}^{\dag}a_{j}
                        + \hbar\sigma_{-}\sum_{j}\ \lambda_{j}a_{j}^{\dag} + \hbar\sigma_{+}\sum_{j}\ \lambda_{j}a_j,
\label{eq:measurement_H}
\]
where $\nu_{a}$ is the transmon's lowest transition energy, $\lambda_{j}$ is qubit-resonator coupling strength, $\sigma_{z}$ is the Pauli
$Z$ matrix and $\sigma_{+}$~($\sigma_{-})$ are the qubit raising (lowering) operator.

The preparation and measurement protocol (Fig. \ref{nmeas}b) would rely upon utilizing the resonant limit of this model. To proceed, the transmon would initially be detuned in energy from all of the modes in the array and prepared in its first excited state with a microwave pi-pulse applied through a coplanar waveguide (CPW) cavity~\cite{blais2004cavity}. Through the application of a flux pulse, the transmon would then be brought into resonance with one of the nanoresonators and allowed to interact for one-half a Rabi cycle ($t_{\mathrm{Rabi}}=\pi/2\lambda_j$), thus transferring its excitation to the nanoresonator mode.  Next, after a predetermined delay, the location of the phonon in the nanoresonator array would be measured by scanning the transmon's  transition energy $\hbar \nu_a$ (via a flux ramp) through the range of nanoresonator energies. Upon achieving resonance with the populated nanoresonator mode, the transmon would transition through a Rabi transfer back to its first excited state, which could be measured through dispersive read-out of the transmon via the CPW cavity~\cite{blais2004cavity,wallraff2005approaching}. As me mentioned the precise location of the nanoresonator could be determined from a map of frequencies using electron beam imaging and finite element simulations. Also, it should be noted that the coherent exchange of excitations through the resonant interaction between a piezoelectric disk resonator and superconducting phase qubit has been demonstrated previously~\cite{ground_state1}; thus it is expected that a similar technique could be adapted to the case considered here.    

For realization of the protocol, and to insure low thermal occupation ($\overline{n}<10^{-1}$) of the nanoresonator modes at dilution refrigerator temperatures, it would be necessary for the nanoresonator array to be composed of ultra-high frequency (UHF) modes in the GHz regime~\cite{huang2003nanoelectromechanical}.  With proper design of the nanostructures'  geometries, the third-order, in-plane flexural modes could be engineered to have frequencies varying over the range of 2 to 4 GHz – note that varying levels of disorder $\Delta$ could be programmed into the array in a controlled manner by deliberating varying the dimensions of the nanostructures.  We remark that the larger the disorder the smaller will the the required array for the observation of localization. However that cannot be freely varied, since a large disorder could mean that the resonators are far from resonance to each other. That would imply the necessity to consider the counterrotating terms in Hamiltonian (3). Therefore is good to keep the limit $J\ll\Delta\ll\overline{\omega_j}$ for the envisaged platform.

We estimate that transmon-nanoresonator coupling strengths $\lambda_{j} \geq$ 1 MHz can be achieved for this configuration by designing a tightly packed array that minimizes the spacing $d$ between nanoresonators (Fig. 1). Table 1 provides estimates of $\lambda_{j}$ (and $J$) for geometrical parameters of the array that are realizable using standard nanolithographic techniques.  With $\lambda/2\pi \geq$ 1 MHz, Rabi transfers between the transmon and nanoresonators would occur over a time-scale of at most 250 ns, which would set the minimum dwell time for each step in the flux scan to determine which nanoresonator mode is populated.  For an array size of 50 nanoresonators, this would yield a maximum scan time of 25 micro-seconds, which in turn would set the minimum tolerable nanoresonator relaxation time.  For GHz-range modes, this would require quality factors in excess of 100,000, which has been achieved with aluminum~\cite{palomaki2013entangling} and carbon-nanotube-based resonators~\cite{moser2014nanotube} at milli-Kelvin temperatures, albeit at much lower resonance frequencies (10s MHz).  
\section*{Tables}
\begin{table}[h!]
\caption{Estimates of $\omega$ assuming the standard expression from
thin-beam theory for the third in-plane flexural frequency with clamped-clamped
boundary conditions. The nanoresonator is assumed to be aluminum, with the
following dimensions: length $L=0.7 \;\mu m \;(0.6 \;\mu m)$ for the $2.5$ GHz
($3.5$ GHz) mode; width $w=45 nm$; thickness $t=50 nm$; The transmon-NR
coupling $\lambda$ was calculated from circuit theory, assuming the systems
to be in resonance, and is given by
$\lambda = \omega \sqrt{\frac{dC}{dx}\frac{C \Delta V ^2}{m \omega^2 d C_Q}}$.
Here $C=20 \; aF, \; dC/dx = 6 \times 10^{-11} F/m$ are the coupling capacitance and
its first derivative respectively, which are estimated from finite element
simulations assuming a gap of $d=20 \;nm$. $C_Q=50 \;fF$ is the transmon shunt
capacitance, which would yield a charging energy of $E_C/h = 400$ MHz. Finally,
$m=0.52 \;.\; \rho w t L$ is the effective mass of the third mode, where $\rho$
is the density of aluminum. The same parameter values are assumed for the
calculation of $J$.}
\begin{tabular}{ | c | c | c | c | }
\hline
 $\omega/2\pi$ (GHz) & $\Delta V$ (V) & $\lambda/2\pi$ (MHz) & $J/2\pi$ (MHz)\\
\hline
 2.5 & 10 & 1.2 & 0.7 \\
\hline
 2.5 & 20 & 2.3 & 2.7 \\
\hline
 3.5 & 10 & 1.2 & 0.6 \\
\hline 
 3.5 & 20 & 2.5 & 2.3 \\
\hline
\end{tabular}
\end{table}\label{tab}

\section{Conclusions and Perspectives}
In this paper, we have devised a quantum simulator based on nanoelectromechanical systems, for analyzing the
many-body effects in quantum systems. Actually, a one-dimensional array of electrostatically coupled nanoresonators is
suggested to simulate the Anderson Hamiltonian. A method is present for coupling the nanoresonator electrostatically, but other means could also be explored to establish larger coupling (such as mechanical links).
By introducing a controlled source of disorder to the system, we studied the localization phenomena in an array of $50$
resonators. Accordingly, the population of the first excited state of a given resonator was analyzed, however, due to the
discrete nature of our system, one could not expect a smooth Gaussian-to-exponential transition in the population profile.

We also studied the influence of thermal effects due to a surrounding environment, arising in real implementation of the
system. By coupling the system to bosonic thermal baths, we studied the interplay between disorder and thermalization.
For sufficiently low temperatures, so that the localization is experimentally detectable with the proposed simulator, the loss of phonons due to the dissipation does not immediately destroy the phonon population localization.
For higher temperatures the localization is affected by the thermal pumping of excitations into the array, which generates
a fully thermalized state.

Initializing the system in a known state and measuring the evolved system, effectively, are important steps in realizing
a quantum simulator. Having coupled the chain of resonators to a superconducting transmon qubit, we have suggested detailed protocols to initialize and measure the system.
By applying a flux pulse, the transmon qubit is brought into resonance with the desired nanoresonator which allows to interact
with the resonator hence transferring its excitation to the nanoresonator.
After system evolution, the location of the phonon in the nanoresonator array can be measured by scanning the transmon's
transition energy through the range of nanoresonator energies. Having achieved resonance with the populated nanoresonator, 
the transmon would transfer back to its first excited state. The transmon can then be measured through dispersive read-out
via the CPW cavity.

Besides simulating Anderson localization, we have also discussed the possibility of using the proposed quantum simulator for implementing the continuous-time quantum walk dynamics. The system allows to realize localization and decoherence in continuous-time quantum walks. The initialization and measurement protocols permit to inject several phonons in the chain to investigate the multi-particle quantum walks. A (non-trivial) two-dimensional version of the suggested simulator can be used to implement two-dimensional quantum walks. In such case, quantum algorithms can be implemented. 
It is worth to mention that to some extent the present discussion applies as well to optomechanical systems, such as in Refs. \cite{Croy,Xuereb},
 whose architectures are worth to be explored for quantum simulation purposes. 
Moreover, given the ability to control individual resonator excitations by scanning the transmon frequency, it is possible to extend the present investigation to the situation including on-site interactions. 
When several realizations are taken into account the effective Hamiltonian averages out to (\ref{eq:A_H}) plus on-site interactions on all sites, in a similar way to the Bose-Hubbard model, allowing investigation of the allowed phases and respective quantum phase transitions when the parameters are varied \cite{bh1,bh2}. 
In that way this would allow for investigation of nonequilibrium steady states of quantum many-body models in similar fashion to other proposed simulators \cite{Angelakis2012,Koch2013}.

\begin{backmatter}

\section*{Competing interests}
  The authors declare that they have no competing interests.

\section*{Author's contributions}
    All Authors have contributed equally to the manuscript.
    
\section*{Acknowledgments}
JKM acknowledges financial support from Brazilian National Council for Scientific and Technological Development
(CNPq), grant PDJ 165941/2014-6. MCO acknowledges support by FAPESP and CNPq through the National Institute for Science and
Technology on Quantum Information and the Research Center in Optics and Photonics (CePOF). MDL acknowledges support for this work provided by the National Science Foundation under Grant DMR-1056423 and Grant DMR-1312421.

\bibliographystyle{bmc-mathphys}
\bibliography{NemChainMLadd}

\end{backmatter}
\end{document}